\documentclass{aa}
\usepackage{psfig}
\usepackage{epsfig}
\def \sax {BeppoSAX}
\def \degmark{^\circ}
\def \nh {N${\rm _H}$}

\def \hcm {\hbox {\ifmmode $ atom cm$^{-2}\else atom cm$^{-2}$\fi}}
\def \arcmin {\hbox{$^\prime$}}
\def \arcsec {\hbox{$^{\prime\prime}$}}

\def\approxgt{\mathrel{\hbox{\rlap{\lower.55ex \hbox {$\sim$}}
        \kern-.3em \raise.4ex \hbox{$>$}}}}
\def\approxlt{\mathrel{\hbox{\rlap{\lower.55ex \hbox {$\sim$}}
        \kern-.3em \raise.4ex \hbox{$<$}}}}
\newcommand{\mc}{\multicolumn}
\begin{document}

\thesaurus{(08; 09.01.2; 09.02.1; 09.07.1; 12.04.2; 13.25.4)}

\title{The low-energy cosmic X-ray background spectrum observed by 
the \sax\ LECS}

\author{A.N. Parmar\inst{1}
        \and M. Guainazzi\inst{1}
        \and T. Oosterbroek\inst{1}
        \and A. Orr\inst{1}
        \and F. Favata\inst{1}
        \and D. Lumb\inst{1}
        \and A. Malizia\inst{2,3}
}
\offprints{A.N.Parmar (aparmar@astro.estec. esa.nl)}

\institute{
        Astrophysics Division, Space Science Department of ESA, ESTEC,
              Postbus 299, 2200 AG Noordwijk, The Netherlands
\and
        BeppoSAX Science Data Center, Nuova Telespazio, via Corcolle 19
        I-00131 Roma, Italy
\and
        Department of Physics and Astronomy, Southampton University,
        SO17~1BJ, UK
}
\date{Received 1998 December 9; Accepted 1999 March 5}

\maketitle

\markboth{LECS background spectrum}{LECS background spectrum}

\begin{abstract}
  
  The spectrum of the 0.1--7.0~keV cosmic X-ray background (CXB) at galactic
  latitudes $>$$\vert 25 \vert \degmark$ has been measured using the
  \sax\ Low-Energy Spectrometer Concentrator (LECS). Above 1~keV the
  spectrum is consistent with a power-law of photon index 
  $1.47 \pm 0.10$ and normalization 
  $11.0 \pm 0.8$~photon~cm$^{-2}$~s$^{-1}$~keV$^{-1}$~sr$^{-1}$ 
  at 1~keV. The overall 
  spectrum can be modeled by a power-law with 2
  thermal components, or by a broken power-law and a single thermal
  component. In both cases the softer thermal emission dominates
  $\approxlt$0.3~keV and is seen through a column, \nh, of a few
  $10^{19}$~atom~cm$^{-2}$. The other components have \nh\ consistent
  with the mean line of sight value. The metal abundances
  for the thermal components are poorly constrained, but consistent
  with cosmic values.  The power-law together with 2 thermal
  components model has been used to fit
  recent combined ASCA and ROSAT CXB measurements. Here,
  the soft thermal component is interpreted as emission from
  the local hot bubble and the hard thermal component as emission from a 
  more distant absorbed region.  While such a 2 component thermal 
  model is consistent with the LECS spectrum, it is not \emph{required},
  and the hard thermal component
  may result from inadequate modeling of the extragalactic contribution.
  The observed
  low-energy spectral complexity may therefore
  originate primarily in the local hot
  bubble. There is no evidence 
  for the presence of a very soft CXB component 
  with a temperature $\approxlt$0.1~keV. The emission measure
  seen by ROSAT is rejected at 90\% confidence.

\end{abstract}


\section{Introduction}

The diffuse Cosmic X-ray Background (CXB) was discovered
in 1962 predating the discovery of the Cosmic Microwave
Background (CMB) by several years. The spectral characteristics and the
spatial distribution of the CXB have been measured by many X-ray
missions, but its origin is still not yet fully understood.  The 
3--60~keV spectrum is well described by a thermal 
bremsstrahlung model with a temperature, kT, of 
40~keV (\cite{marshall80}).  Between 1 and 10~keV this
shape can be approximated by a power-law with a photon index,
$\alpha$, of 1.4 (\cite{jahoda92}; \cite{gendreau95}). The lack of
distortion of the CMB expected from a truely diffuse hot 
intergalactic medium 
supports the idea that the CXB above 1~keV is
dominated by emission from an unresolved distribution of
active galactic nuclei (AGN; \cite{setti89}; \cite{madau93};
\cite{comastri95}).  Recent results from the deepest (flux limit
$\sim$10$^{-15}$~erg~cm$^{-2}$~s$^{-1}$) X-ray survey so far
performed, suggest that the fraction due to discrete sources in the
energy range 0.5--2.0~keV is 70--80\% (\cite{hasinger98}).
Recent results from a \sax\ survey support such an origin
(\cite{fiore98}).

Below $\sim$1.5~keV the CXB shows additional spectral
and spatial complexity. Spatially, there
is a large degree of structure visible, with
several features evident in all-sky maps (e.g, \cite{mccammon90};
Snowden et al. 1995). At $\sim$0.25~keV there is
a clear anti-correlation with both galactic latitude (the CXB being
significantly weaker at the galactic equator), and with the amount
of H\,{\sc i} in the line of sight
(\cite{snowden94}). The low-energy CXB spectrum 
deviates from the simple power-law observed $\approxgt$1~keV.
Line emission from O\,{\sc vii} and C\,{\sc v} was
observed using solid-state detectors by \cite{rocchia84} and
\cite{inoue80}, strongly
supporting the view that emission from an optically thin thermal 
plasma contributes significantly at low-energies.  Later studies (e.g.,
\cite{kerp93}) show that to fit  
ROSAT Position Sensitive Proportional Counter (PSPC)
spectra of the soft CXB two thermal components are required, in
addition to the power-law.

The spatial dependence of the low-energy CXB is similar to that
expected if extragalactic, isotropic emission is absorbed by
the galactic interstellar medium (ISM). However, the observed degree
of modulation of the CXB and the lack of a significant spectral
modulation (\cite{mccammon83}) are incompatible with such a model.
Thus, a significant part of the soft CXB emission is likely to be
of local origin. Models which rely on a
significantly clumpy ISM (\cite{jakobsen86}) 
have been proposed to explain the lower
than expected modulation from an isotropic ISM. 
Diffuse emission maps and deeper pointed observations of shadowing by
cold gas (\cite{snowden97}; \cite{kerp94}; \cite{kerp99})
have revealed a consistent yet complex origin for the 
diffuse background. A soft ($\sim$0.1~keV) component originates from a
combination of the local hot bubble and a galactic halo, but with the 
latter modulated spatially 
by local enhancements from bright clouds. In the 0.75~keV 
energy range the galactic 
emission is dominated by a bulge component and other enhancements
associated 
with relatively nearby supernova remnants and wind-blown bubbles, 
together with a galactic plane enhancement of which some component is 
undoubtedly due to a superposition of individual sources. There may also be
a contribution from inter-cluster gas of the Local Group (see
Hasinger 1996 and references therein).

Deep pointed observations performed with imaging detectors onboard
ROSAT and ASCA have provided a good opportunity to study the 
0.1--10~keV CXB, typically yielding, as discussed above,
a spectrum compatible with a power-law and two thermal components.
However, the spectral parameters derived from the two missions are not
always consistent. In particular, the power-law slopes are not always
in agreement (see the discussion in Chen et al. 1997) and
even in the well calibrated 1--2~keV overlapping energy range, 
estimates of the CXB flux vary by 15--20\%.  
Even measurements using
the same instrument (the PSPC) differ significantly in spectral
parameters (cf.\ \cite{hasinger92} and \cite{georgantopoulos96}).

\section{The instrument and observations}
\label{sect:obs}

The Low-Energy Concentrator Spectrometer (LECS; 0.1--10~keV;
\cite{parmar97}) is an
imaging gas scintillation proportional counter on-board the
Italian-Dutch \sax\ X-ray astronomy mission (\cite{boella97}). 
In comparison with the 0.1--2.5~keV PSPC, the LECS
offers energy resolution a factor $\sim$2.4 better and an extended
high-energy response, overlapping with the ASCA Solid State Imaging
Spectrometer (SIS; 0.5--10~keV). The LECS has comparable energy
resolution to the SIS at $\sim$0.5~keV. This combination of
moderate energy resolution and broad energy coverage provides 
a unique capability for studying the low-energy CXB.
In particular, the determination of the normalization and break energy
of the CXB spectrum around 1~keV has proved problematical. The LECS spans
this energy range exactly and so provides the potential for 
accurately measuring the relative contributions of the
galactic and extragalactic components. 

The LECS has a circular field of view (FOV) of 37$'$ diameter
(\cite{parmar97}) and an effective area of $\sim$10~cm$^2$ at 0.28 keV
and $\sim$40~cm$^2$ at 2~keV.  In the central 8\arcmin\ radius of the
FOV, the mean Non X-ray Background (NXB) 0.1--10~keV counting rate is
$5.2 \times 10^{-6}$~s$^{-1}$~keV$^{-1}$~arcmin$^{-2}$
(\cite{parmar99}).  The 1.25~$\mu$m thick LECS entrance window is
supported by a fine grid and a strongback. The strongback divides the
aperture into $4' \times 4'$ squares, each of which is divided into a
matrix of 8 by 8 squares by the fine grid.  The obscuration caused by
the fine grid is energy independent and does not vary significantly
with FOV position. In contrast, the strongback produces
a position dependent obscuration which is also slightly
energy dependent due primarily to the variation of
the mirror point spread function with energy (\cite{parmar97}).

%----------------------------Table 1-----------------------------------
\begin{table*}
\caption[]{LECS observations used to create the
CXB spectrum. A target name including ``sec'' or 
``secondary'' refers to the prime Wide Field Camera target in a direction
orthogonal to the LECS pointing. N is the number of
individual pointings. ${\rm N_{gal}}$
is the line of sight absorption in units of $10^{20}$~\hcm\ 
(\cite{dickey90})}
\begin{flushleft}
\begin{tabular}{llllllrlrl}
\hline\noalign{\smallskip}
       Target & \mc{2}{c}{Observation} & N & \mc{2}{c}{LECS Pointing (J2000)} 
       & ${\rm l_{II}}$ \hfil
       & \hfil ${\rm b_{II}}$ & Exp. & ${\rm N_{gal}}$\\
       & \hfil Start  & \hfil Stop&& \hfil RA & \hfil Dec & $(\degmark)$ \hfil 
       & \hfil $(\degmark)$ & (ks) & \\
       & (yr~mn~day) & (yr~mn~day) & & \hfil (h~m~s) & \hfil ($\degmark\ 
       \arcmin\ \arcsec$) & & & & \\
\noalign{\smallskip\hrule\smallskip}
Polaris Region   & 1996 Jul 01 & 1997 May 23 & 6 &02 31 42.0 & +89 15 47 & 
123.3 & +26.5 & 69.7 & 7.0 \\
%Cen X-3 (sec)   & 1996 Aug 11 & 1996 Aug 14 & 1 &14 42 27.1 & +19 44 10 & 
%22.7 & +63.6 & 22.1 & 2.3 \\
Gal cent-2 (sec) & 1996 Aug 23 & 1996 Aug 27 & 1 &16 51 19.5 & +60 11 49 & 
89.8 & +38.1 & 82.3 & 2.1\\
Gal cent-3 (sec) & 1996 Aug 27 & 1996 Aug 31 & 1 &16 35 10.7 & +59 46 30 & 
89.9 & +40.2 & 79.9 & 2.0\\
SGR-A (sec)      & 1996 Oct 10 & 1996 Oct 12 & 1 &06 12 22.7 & $-$60 59 04 & 
270.1 & $-$28.2 & 43.3 & 4.2\\
Secondary Target (S1)& 1997 Mar 18 & 1997 Mar 25 & 1 &17 56 46.1 & +61 11 45 & 
90.2 & +30.1 & 94.2 & 3.4\\
Secondary Target (S2)& 1997 Apr 13 & 1997 Apr 15 & 2 &18 18 20.5 & +60 58 42 & 
90.2 & +27.4 & 26.6 & 3.8\\
SDC Target       & 1997 Dec 13 & 1997 Dec 14 & 1 &23 07 53.5 & +08 50 06 &
84.4 & $-$46.1 & 13.9 & 4.7 \\
Secondary Target (S4)& 1998 Mar 10 & 1998 Mar 22 & 2 &05 52 07.9 
& $-$61 05 35 & 270.0 & $-$30.6 & 46.7 & 5.3 \\
Secondary Target (S5)& 1998 Aug 22 & 1998 Oct 01 & 4 &17 52 07.4 & +61 01 01 &
89.9  & +30.6 & 102.2 & 3.5 \\
\noalign{\smallskip}
\hline
\end{tabular}
\end{flushleft}
\label{tab1}
\end{table*}

%----------------------------Table 1-----------------------------------

A 559~ks exposure background spectrum was accumulated from a number of
high galactic latitude exposures by extracting counts within an
8\arcmin\ radius of the nominal source position within the LECS FOV
(Table~\ref{tab1}). The choice of background fields and LECS background
subtraction methods are discussed in Parmar et al. (1999).
The background fields were selected to avoid any of the obvious brightness
enhancements in the ROSAT maps in order to provide a reliable
estimate of the spectrum at ``typical'' locations out of the galactic
plane. The CXB spectrum is expected to include contributions from 
emission in the local hot bubble, the galactic halo, and unresolved AGN.

No point sources are present in the individual fields with 0.1--2.0~keV
fluxes $>$$1.7 \times 10^{-13}$~erg~cm$^{-2}$~s$^{-1}$.
The background spectrum includes contributions from the CXB, the NXB (which is
internal to the instrument), and any X-rays from within a solid angle
of 2$\degmark$ which may undergo a single reflection and be detected
in the FOV (\cite{conti94}).  The extraction region was chosen to
include 95\% of the 0.28~keV X-rays and is the region where the
instrument is best calibrated.  All the fields have galactic latitudes
of $>$$\vert 25\degmark \vert$ and Galactic column densities of
between 2.0 and $7.0\times 10^{20}$~\hcm.  

The contribution
of the NXB was subtracted using a 499~ks NXB spectrum accumulated 
during intervals of dark Earth pointing.
The properties of this spectrum are discussed in Parmar et al. (1999).
Briefly, in the central 8\arcmin\ the LECS NXB spectrum is 
approximately constant with energy with a count rate of 
$8 \times 10^{-4}$~s$^{-1}$~keV~$^{-1}$. 
Above $\sim$4~keV there is a
smooth increase in count rate with 3 discrete line features 
superposed. In the energy ranges 0.1--0.5~keV, 0.5--2.0~keV and
2.0--5.0~keV the NXB contributes 13\%, 19\%, and 47\% of the
total background, respectively. Above $\sim$8~keV,
the NXB dominates the overall background spectrum.
The LECS NXB shows a gradual reduction in counting rate by $\sim$15\%
over an interval of $\sim$2~years (\cite{parmar99}). Due to the
low-inclination, almost circular, \sax\ orbit,
variations in the NXB counting rate around the orbit are negligible.

All data products were accumulated using the latest version of the
LECS analysis software ({\sc saxledas 1.8.0}).  The spectrum was
rebinned to have at least 20 counts per bin and to sample the
full width half-maximum of the energy resolution with at most 3 energy bins, in
order to ensure the applicability of the $\chi ^2$ statistic. All
uncertainties are quoted at the 90\% confidence level for one
interesting parameter ($\Delta \chi^2 = 2.7$). When determining
spectral uncertainties all the other variable parameters were allowed 
to vary freely. 
The photoelectric
absorption coefficients of \cite{morisson83}, together with the cosmic
abundances of \cite{anders89} were used (which are supposed to be
representative of the composition of the proto-solar nebula).  A
response matrix appropriate for uniform diffuse emission was created
using {\sc lemat 3.5.1}. This matrix differs from the standard
point-source response matrix in the following ways:

\begin{enumerate}
  
\item{} The nominal target position lies close to the center of a window
  support strongback square and the on-axis transmission is a factor 1.08 
  higher than the average for the entire extraction region.
  
\item{} The effects of off-axis mirror vignetting are included.

\item{} The CXB is modeled as a uniform diffuse emission,
  by means of 100 point sources randomly distributed within a radius
  of 12\arcmin. This radius is larger than that of the extraction region
  (8\arcmin) to allow for photons from sources located within the 
  region that are spread outside of the region 
  by the finite detector resolution, and for 
  events originating outside the extraction region which are
  spread inside.

\end{enumerate}

In addition to the above effects, 
single reflected X-rays from within 120\arcmin\ can
be detected in the FOV (\cite{conti94}). In order to estimate
the magnitude of this effect, 0.1--6~keV flux
ratios between a pointed Crab Nebula observation (\cite{cusumano99}) and two
pointings where the Crab Nebula was located just outside the LECS FOV,
with offset angles of 45\arcmin\ and 60\arcmin, 
were calculated.
The spectrum of the singly reflected X-rays is consistent with that
observed on-axis, but with a much reduced intensity.
An exponential function was used to characterize the dependence
of the flux ratios on offset angle. Assuming that the
X-ray sky is uniform within 120\arcmin, integration of
this function reveals that singly scattered X-rays contribute $<$1\%
of the flux within an 8\arcmin\ extraction radius. Since this is well
within the uncertainty in CXB normalization, this effect is ignored.

\section{X-ray background spectrum}
\label{sect:bgfits}

\begin{figure}
\centerline{
      \hbox{
      \psfig{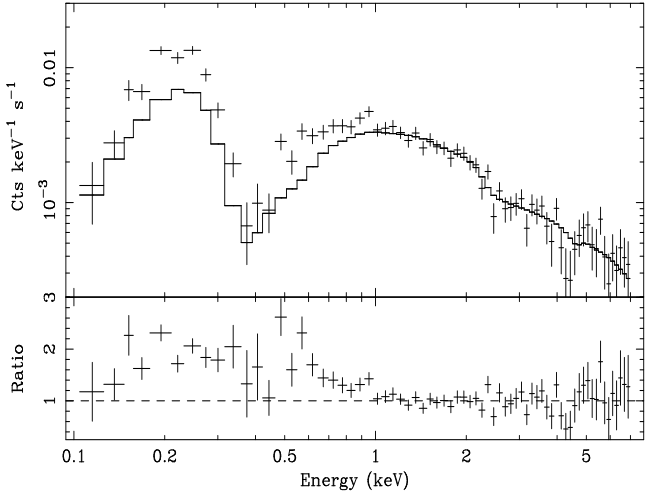}
}}
\caption{Spectrum and residuals in units
of the data/model ratio (lower panel), when the best-fit 1--7~keV 
power-law is extrapolated down to 0.1~keV}
\label{fig1}
\end{figure}

In the 1.0--7.0~keV energy range (at higher energies there is no
signal at $>$3$\sigma$ and the LECS spectral calibration is more uncertain)
a simple power-law fit gives an acceptable $\chi^2$ of 39.0
for 44 degrees of freedom (dof) with $\alpha = 1.47
\pm 0.10$ and a normalization of
$11.0 \pm 0.8$~photon~s$^{-1}$~cm$^{-2}$~keV$^{-1}$~sr$^{-1}$ at
1~keV.
Significant excess counts are present in the spectrum at lower
energies (see Fig.~\ref{fig1}).

%-----------------------------Figure 1--------------------------------
%-----------------------------Figure 1--------------------------------

A good fit can be obtained in the 0.1--7.0~keV energy range with a low
absorption (${\rm N_H = 5 \times 10^{19}}$~atom~cm$^{-2}$) broken
power-law, provided that a narrow (i.e.\ intrinsic width
$\sigma$ fixed at 0) Gaussian emission line is added to the
continuum (model~1).  The centroid energy implies that this
feature is 0.57~keV K\,$\alpha$ emission from O~{\sc vii},
indicative of plasma at a temperature of
$\sim$0.1~keV. Best-fit parameters are listed in
Table~\ref{tab2}. 

%-----------------------------Table 2--------------------------------
\begin{table}
\caption{Best-fit parameters for model~1 consisting of a
    broken power-law
    with a photon index ${\rm \alpha_{soft}}$ below energy ${\rm
    E_{break}}$ and ${\rm \alpha_{hard}}$ above. A
    narrow Gaussian line at an energy ${\rm E_{line}}$
    with normalization ${\rm I_{line}}$ is included}
\begin{tabular}{lc}
\hline\noalign{\smallskip}
Parameter & Value \\
\noalign {\smallskip}
\hline\noalign {\smallskip}
Broken power-law                              & \\
\quad \nh\ ($10^{20}$~atom~cm$^{-2}$)         & $0.5 \pm ^{0.4} _{0.3}$ \\
\quad ${\rm \alpha_{soft}}$                   & $2.03 \pm 0.14$ \\
\quad ${\rm E_{break}}$ (keV)                 & $1.4 \pm _{0.2} ^{0.3}$ \\
\quad ${\rm \alpha_{hard}}$                   & $1.43 \pm 0.13$ \\
Gaussian line                                 & \\
\quad ${\rm E_{line}}$ (keV)                  & $0.57 \pm 0.05$ \\
\quad ${\rm I_{line}}$ (ph~cm$^{-2}$~s$^{-1}$) & 
$(6 \pm 4) \times 10^{-5}$ \\
${\rm \chi^2/}$~dof                       & 69.5/63 \\
\noalign {\smallskip}
\hline
\end{tabular}
\label{tab2}
\end{table}
%-----------------------------Table 2--------------------------------

\begin{figure}
\centerline{
      \hbox{
      \psfig{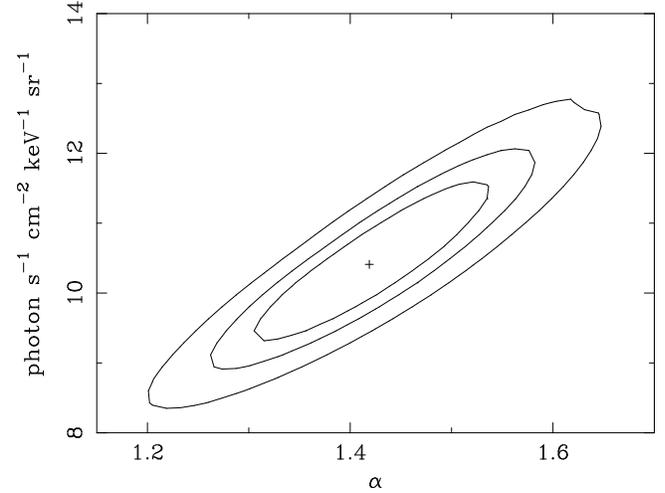}
}}
\caption{Iso-$\chi^2$ contours for the power-law normalization versus 
spectral index using model~2. Contours correspond to $\Delta \chi ^2$
= 2.3, 4.6 and 9.2}
\label{figmg7}
\end{figure}

Following the current understanding of the CXB spectrum,
the 0.1--7~keV LECS spectrum was next fit with a
power-law together together with a single temperature optically thin plasma
(the Mewe-Kaastra-Liehdal plasma emissivity model
in {\sc xspec}, \cite{mewe85}, 1986).
This gives an unacceptable fit at $>$95\% confidence
(${\rm \chi^2 = 83}$ for 62~dof).
An acceptable fit is obtained if an absorbed power-law is used,
together with two thermal components with cosmic abundances
(model~2). The \nh\ of the low temperature thermal component
was a free parameter in the fit, while the \nh\ of the other thermal
and the power-law components was held fixed 
at the exposure weighted average of the galactic
columns in the directions of each of the pointings listed in
Table~\ref{tab1} (${\rm 3.7 \times 10^{20}}$~atom~cm$^{-2}$). 
The $\chi^2$ is 66.5 for 63 dof (null hypothesis probability
36\%). The addition of the second thermal
component provides an improvement in fit quality 
at $>$99.9\% confidence (${\rm
\Delta \chi^2 = 13.6}$ for 2 dof).  Fig.~\ref{figmg7} shows a
%-----------------------------Figure 7--------------------------------
%-----------------------------Figure 7--------------------------------
contour plot for the power-law spectral index versus normalization.
The best-fit temperatures are
$0.7 \pm _{0.3} ^{0.2}$~keV and $0.137 \pm^{0.011}_{0.022}$~keV for the high
and low absorption components, respectively (see Table~\ref{tab3}).

%-----------------------------Table 3--------------------------------
\begin{table}
\caption{Best-fit parameters for model~2 consisting of a power-law
  and two thermal plasmas. Abundances are fixed at cosmic values. 
  The absorption of the power-law and hard thermal components 
  was fixed at the mean line of sight
  value. The values in square brackets are the one-sided 
  variations in best-fit values when \nh\
  is fixed at 
  2.0 and $7.0\times 10^{20}$~atom~cm$^{-2}$. The power-law normalization
  is in units of photon~cm$^{-2}$~s$^{-1}$~keV$^{-1}$~sr$^{-1}$ at 1~keV} 

\begin{tabular}{lc}
\hline\noalign{\smallskip}
Parameter & Value \\
\noalign {\smallskip}
\hline\noalign {\smallskip}
Power-law                              & \\
\quad \nh\ ($10^{20}$~atom~cm$^{-2}$)  & 3.7 \\ 
\quad $\alpha$                         & $1.42 \pm 0.10$ [0.02] \\
\quad Normalization                    & $10.4 \pm ^{1.4} _{1.1}$ 
[0.3] \\
``Soft'' thermal                       & \\
\quad kT (keV)                         & $0.137\pm^{0.011}_{0.022}$ [0.002] \\
\quad Normalization (ph~cm$^{-5}$) & $(2.6  \pm^{1.4}_{0.6}) \times 10^{-4}$ [0.3] \\ 
\quad Abundance                        & 0.04--9.5 \\
\quad \nh\ ($10^{20}$~atom~cm$^{-2}$)  & $0.7 \pm ^{0.4} _{0.3}$ [0.1]  \\ 
``Hard'' thermal                       & \\
\quad kT (keV)                         & $0.7 \pm ^{0.2} _{0.3}$ [0.03] \\
\quad Normalization (ph~cm$^{-5}$) & $(3.1 \pm 1.4) 
\times 10^{-5}$ [0.5] \\
\quad \nh\ ($10^{20}$~atom~cm$^{-2}$)  & 3.7 \\ 
${\rm \chi^2/}$~dof                    & 66.5/63 \\
\noalign {\smallskip}
\hline\noalign {\smallskip}
\end{tabular}
\label{tab3}
\end{table}
%-----------------------------Table 4--------------------------------
\begin{table}
\caption{Best-fit parameters for model~3 consisting of a
  broken power-law and a single thermal component.
  The abundance is fixed at cosmic values. The
  broken power-law \nh\ was fixed at the mean line of sight
  value and its normalization
  has units of photon~cm$^{-2}$~s$^{-1}$~keV$^{-1}$~sr$^{-1}$ at 1~keV}
\begin{tabular}{lc}
\hline\noalign{\smallskip}
Parameter & Value \\
\noalign {\smallskip}
\hline\noalign {\smallskip}
Broken power-law                              & \\
\quad \nh\ ($10^{20}$~atom~cm$^{-2}$)  & 3.7 \\ 
\quad ${\rm \alpha_{soft}}$                         & $2.1\pm ^{0.3} _{0.2}$ \\
\quad ${\rm \alpha_{hard}}$                         & $1.46\pm 0.14$  \\
\quad Break energy (keV)                & $1.4 \pm ^{0.4} _{0.2}$ \\
\quad Normalization &  $12.7 \pm 0.6$ \\
Thermal                       & \\
\quad kT (keV)                         & $0.12\pm ^{0.02} _{0.03}$ \\
\quad Normalization (ph~cm$^{-5}$) & $(2.0\pm ^{1.3} _{0.7}) \times 10^{-4}$ \\
\quad \nh\ ($10^{20}$~atom~cm$^{-2}$)  & $0.5 \pm ^{0.6} _{0.5}$ \\ 
${\rm \chi^2/}$~dof                    & 70.0/63 \\
\noalign {\smallskip}
\hline\noalign {\smallskip}
\end{tabular}
\label{tab4}
\end{table}
%-----------------------------Table 4--------------------------------

%-----------------------------Figure 5/6--------------------------------
\begin{figure*}
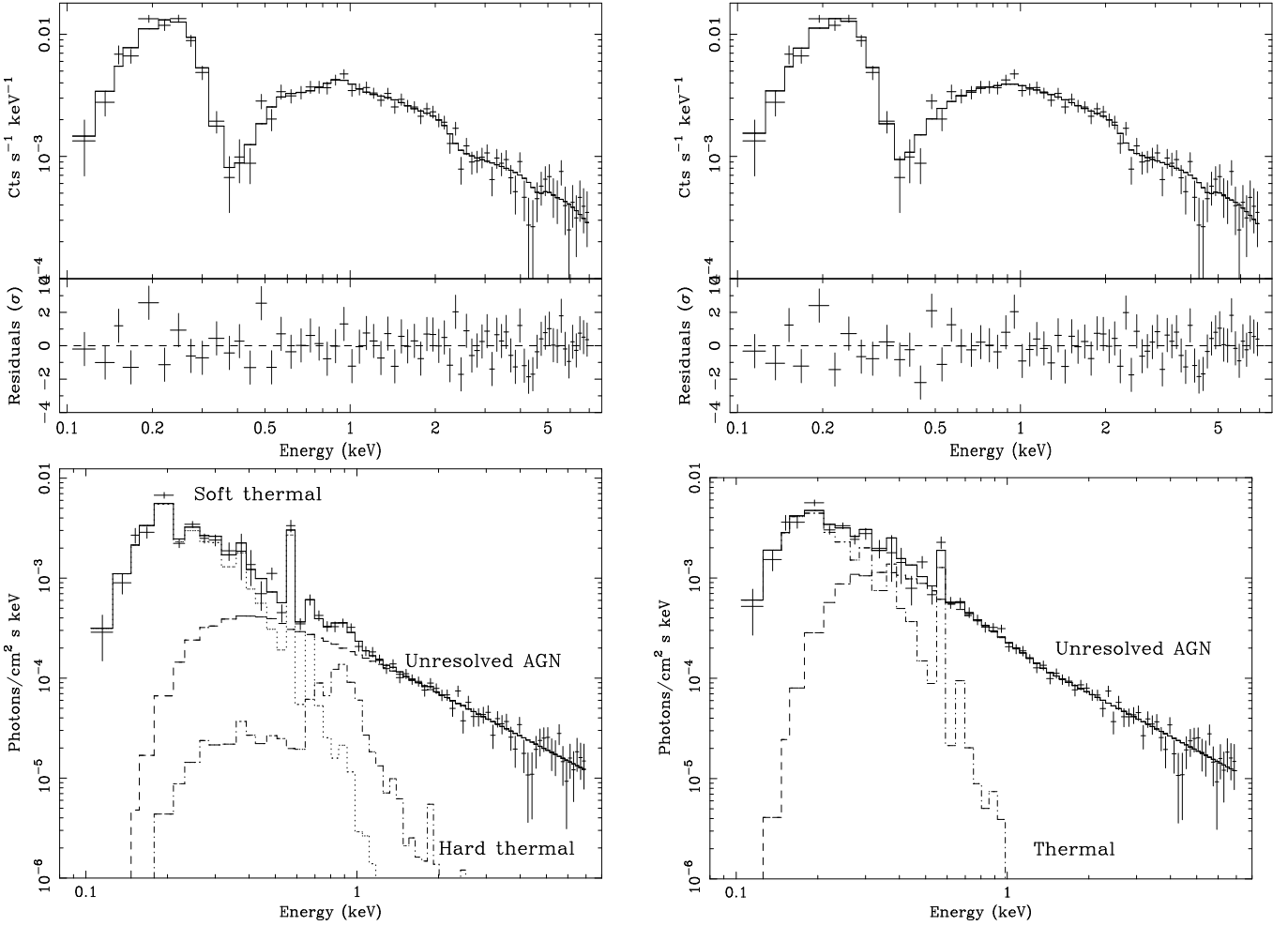

      \epsfig{figure=h1311f3a.eps,height=8.5cm,width=6.5cm,angle=-90}
      \hspace{0.5cm}
      \epsfig{figure=h1311f3b.eps,height=8.5cm,width=6.5cm,angle=-90}
      \epsfig{figure=h1311f3c.eps,height=8.5cm,width=6.5cm,angle=-90}
      \hspace{0.5cm}
      \epsfig{figure=h1311f3d.eps,height=8.5cm,width=6.5cm,angle=-90}
\caption{The left panels show the power-law and two thermal model 
        fit (model~2; Table~\ref{tab3}). The right panels show the 
        broken power-law and 
        single thermal model fit (model~3; Table~\ref{tab4}).
        Observed spectra are shown in the upper panels, fit residuals in
        units of standard deviations in the middle panels and the unfolded
        spectra in the lower panels}
\label{figmg5}
\end{figure*}
%-----------------------------Figure 5/6--------------------------------

The softer thermal component is observed through an absorbing column
of $7 \times 10^{19}$~atom~cm$^{-2}$, consistent with the
emission originating from $\sim$50~pc (assuming
a mean galactic plane local density of 0.4~atom~cm$^{-3}$, \cite{bloemen87}).
Fig.~\ref{figmg5} shows the observed
spectrum and best-fit model obtained using the parameters listed in
Table~\ref{tab3}.  The soft component is
responsible for the majority of the oxygen line emission evident in the
spectrum. The 1--2~keV CXB flux is $1.47 \times
10^{-8}$~erg~cm$^{-2}$~s$^{-1}~$sr$^{-1}$.  The ratio between the
fluxes of the thermal and non-thermal components in
the 0.5--2.0~keV energy range is 0.55.

The quality of the data is not sufficient to strongly constrain the
abundances of the plasmas responsible for the thermal emission.
Leaving the metal abundances free to vary does not result in a
significant improvement in fit quality. In particular, the
abundance of the hotter component is essentially unconstrained. For
the soft component the abundance can be constrained to be between
0.04 and 9.5 times cosmic values.

The summing of observations from regions of the sky
with different amount of absorbing column can create
artificial features in the spectrum, which may result in an incorrect
estimation of spectral parameters. Therefore, the sensitivity of
the derived spectral parameters to the assumed absorbing column
density was investigated. 
${\rm N_H}$ was fixed at the lowest and highest
line-of-sight ${\rm N_H}$ values in
Table~\ref{tab1} of 2.0 and $7.0\times 10^{20}$~atom~cm$^{-2}$
and the fit repeated. 
The changes in best-fit parameters are within their statistical
uncertainties and shown in square brackets in
Table~\ref{tab3}. The same model was also applied to the spectra
from the individual pointings. The temperature of the ``hard''
component is not well constrained, 
and was therefore fixed at
its best-fit value given in Table~\ref{tab3}.
Again, none of the best-fit parameters
differ from their average values by more than 1$\sigma$.
This analysis indicates that the best-fit spectral results are
insensitive to the range of likely absorbing column densities.

An alternative spectral model, requiring only one soft thermal
component (model~3) fits the observed background spectrum at 
a similar
level of confidence. If most (or all) of the power-law component of
the CXB is due to unresolved AGNs, the use of a single
power-law spectral model across such a wide energy range may be incorrect.
In the energy range 2--10~keV the average spectra of AGN may
be represented by a power-law with the ``canonical'' observed $\alpha$ of 
$\sim$1.7 (\cite{mushotzky84}; \cite{nandra94}). 
Below this energy typical AGN spectra steepen 
to $\alpha \sim 2.5$ (\cite{yuan98})
becoming no lower than $\sim$2 at very faint flux levels
(\cite{hasinger93}; \cite{almaini96}). ``Type 2'' (absorbed)
AGN exhibit soft X-ray components, seen either in
transmission or as reflection above the strongly depressed nuclear
radiation (see e.g., \cite{turner97}).
Thus, the simple power-law used in
the modeling is in principle inadequate and should be substituted by 
more complex model, such as a
broken power-law. However, a broken power-law 
together with two thermal components model
gives an insignificant improvement in fit quality to the LECS spectrum
($\Delta \chi^2 = 0.05$ for 2 additional dof).

However, a statistically acceptable fit, of similar quality as the
single power-law plus two thermal components, can be obtained with a
broken power-law model and a single temperature optically thin plasma
giving a $\chi^2$ of 70.0 for 63 dof (model~3, null
hypothesis probability 26\%). 
In this case the more 
absorbed thermal component is not required (the
$\Delta \chi^2$ being $0.7$ for the addition of a hard thermal
component). The best-fit parameters are given in
Table~\ref{tab4}. The low-energy power-law spectral index is ${\rm
\alpha_{soft} = 2.1 \pm ^{0.3} _{0.2}}$, while the parameter values in common
with model~2 are only
marginally different. The ratio between the fluxes of the thermal and
non-thermal components in the 0.5--2.0~keV energy range is 0.14.
The thermal component is 
absorbed with a low \nh, again consistent with an origin
within the local hot bubble.
Neither model~2 nor~3 requires the presence of a
``super-soft'' component at a temperature of $\approxlt$0.1~keV 
which is commonly required
to give an acceptable fit to the PSPC spectra of the soft X-ray CXB
(${\rm \Delta \chi^2}$ = 0.1 for the
addition of 2 dof). 
Miyaji et al. (1998) estimate the normalization and temperature of the
``super-soft'' component from observations of the Lockman Hole and 
Lynx-3A fields.
When the PSPC temperatures of the Lockman Hole are considered,
the LECS 90\% confidence upper limit
normalization is 6.8 in the units used by Miyaji et al. (1998). 
These authors determine a 
normalization of $9.0 \pm ^{2.5} _{1.4}$ (90\% confidence uncertainties). 
For the Lynx-3A temperature range, 
the LECS 90\% confidence upper limit normalization 
is 7.9 and and the PSPC measurement $8.1 \pm 0.3$.

Given that most previous spectral fits to the CXB have
generally used different codes for the emission of optically
thin collisionally ionized plasma (the most common being the
Raymond-Smith code in {\sc xspec}),
the sensitivity of the best-fit parameters to the choice of
spectral emissivity model (Raymond-Smith versus Mewe-Kaastra-Liedhal)
was checked. The variations, at the present signal-to-noise
ratio, are small, with differences in the best-fit temperatures of
$\le$40 and $\le$140~eV for the soft and hard thermal components,
respectively.

\section{Discussion}

The LECS CXB spectrum is in qualitative agreement with results
obtained from PSPC and SIS spectra which are well
represented by a composite spectrum consisting of a power-law 
and two soft thermal components. Both the slope
and the normalization of the power-law component 
are in good agreement with
previous measurements based on ASCA SIS data. The spectral index
derived from the LECS data is $1.47 \pm 0.10$,
while values around
$\simeq$1.4 have been derived from ASCA data
(\cite{chen97}; \cite{miyaji98}). 
The intercalibration of the LECS and ASCA SIS is 
discussed in \cite{orr98}.
The power-law index derived from the harder part of
the CXB spectrum from PSPC data alone is
however significantly different ($1.53 \pm 0.07$) than the value obtained
using SIS data from the same region of sky (\cite{chen97}).
These authors attribute this discrepancy to
``calibration uncertainties in ROSAT''. Clearly, further
comparisons of ROSAT PSPC, ASCA SIS, and BeppoSAX spectral results
are desirable.

There is still no agreement on the normalization of the CXB, and the
comparison is made even more difficult by the different energy bands
over which the measurements have been made.
The 1~keV normalization
of a simple power-law fit above 1~keV in the LECS
($11.0 \pm 0.8$~photon~s$^{-1}$~cm$^{-2}$~keV$^{-1}$~sr$^{-1}$) is
slightly higher than observed by ASCA 
(9--10~photon~s$^{-1}$~cm$^{-2}$~keV$^{-1}$~sr$^{-1}$, \cite{gendreau95}; 
\cite{miyaji98}). This results in a 20\% higher
1--2~keV flux (1.47 
versus $1.22 \times 10^{-8}$~erg~cm$^{-2}$~s$^{-1}$~sr
$^{-1}$).
However, the LECS value is in good agreement with both the ROSAT PSPC
flux ($1.44\times 10^{-8}$~erg~cm$^{-2}$~s$^{-1}$~sr$^{-1}$,
\cite{hasinger96}) and the combined ROSAT/ASCA measurement 
($1.46 \times 10^{-8}$~erg~cm$^{-2}$~s$^{-1}$~sr$^{-1}$, 
\cite{chen97}). We note that 
the 2--6~keV LECS flux is also in
good agreement with that measured independently by the \sax\
MECS (\cite{molendi97}).

The contribution of the power-law component in the 0.5--2.0~keV
energy range is $\simeq$65\%. This is 
close to the ``best guess'' that Hasinger et al. (1998) derive from
the ROSAT deep survey results for the contribution of discrete
sources of 70--80\%.
This fraction rises to 87\% if the LECS CXB spectra is instead fit
with model~3. 
The temperatures of the two thermal components derived from the LECS
spectrum are significantly different from those
obtained with the ROSAT PSPC. The best-fit
LECS temperatures are $0.137 \pm ^{0.011} _{0.022}$~keV and 
$0.7 \pm ^{0.2} _{0.3}$~keV, while temperatures of
$0.06 \pm 0.015$~keV and $0.17 \pm 0.045$~keV (\cite{kerp94}), 
and $\approx$0.07~keV and $\approx$0.14~keV
(\cite{miyaji98}) are obtained with the PSPC.
In addition, the CXB spectra obtained with the low-energy
detectors of the A2 experiment onboard HEAO-1
can also be modeled with two thermal components with temperatures of
0.12~keV and 0.18~keV (\cite{garmire92}).

Thus, while the component with kT $\sim$0.15~keV is seen
by both the LECS and the PSPC (and is close to the average of the
temperatures obtained with HEAO~1), there is
no evidence, in the LECS CXB spectra, for a very soft component with a
temperature $\approxlt$0.1~keV. A similar discrepancy is obtained
when results obtained with the PSPC are compared with those from
other instruments. As an example, \cite{jordan98}
compare the differential emission measures (DEM) obtained from
line fluxes in Extreme Ultraviolet Explorer (EUVE)
spectra of active stars with
those derived by simple two temperature fits to
ROSAT PSPC spectra of the same sources. 
While the emission measures of the hotter 
components appear in good agreement with the EUVE DEM, those of 
the cooler components are a factor $\approxgt$10 higher than the EUVE 
emission measures at the same temperature
(e.g, Figs.~12--14 of Griffiths \& Jordan 1998). These figures
also show that there is no large disagreement between the EUVE DEM 
and emission measures derived from ASCA SIS and {\it Einstein} IPC and 
Solid State Spectrometer spectra.
Similar results are
obtained by \cite{brick98} comparing the PSPC and EUVE spectra of
the active binary 44~Boo. 
The discrepancy between the LECS and
PSPC based analyses of the CXB spectrum is of the same nature, with
the PSPC analysis deriving a cool component with a large emission
measure, not visible in the LECS. This points to a 
possible systematic effect in the low-energy calibration of the
PSPC.

Analysis of the LECS CXB shows that, in addition to the power-law plus
two thermal components, the spectrum can also be explained with a
single thermal component plus a broken power-law.  The best-fit
temperature for the single thermal component of 0.12~keV is very
similar to the temperature of the cooler component obtained in the two
temperature fit. Thus, the effect of introducing the broken power-law
in the fit is to eliminate the 0.7~keV thermal component. 
Additional evidence for the presence of multiple components to the CXB
comes from shadowing experiments using nearby ($\sim$60~pc) clouds
such as LVC~88$+$36$-$2 and MBM~12 (\cite{kerp94}; \cite{snowden93}).
Differences between on- and off-cloud PSPC spectra indicate the
presence of a nearby $\sim$0.06~keV thermal component associated
with the local hot bubble and a more distant absorbed $\sim$0.17~keV 
thermal component located beyond the Galactic H~{\sc i} layer (\cite{kerp94}).
The distant component
dominates the off-cloud low-energy ($<$0.3~keV) spectrum.
In observations towards clouds located within the local hot bubble,
the distant component suffers extra absorption, as does only part
of the nearby component.

It is difficult to directly compare the LECS and PSPC spectral results.
Both instruments indicate that emission from a plasma with a temperature of
$\sim$0.15~keV dominates the low-energy CXB spectrum. In the case
of the PSPC this is the distant, absorbed, component (\cite{kerp94}).
In contrast, the LECS indicates that this is the nearby component
and the more distant (absorbed) component 
is either a 0.7~keV thermal plasma, or the low-energy tail of a broken 
power-law. If this latter model is correct, then there is no
need to invoke a second thermal component. The broken power-law
almost certainly results from distant AGN.
The differences between on- and off-cloud spectra may result primarily
from emission measure and absorption effects within the local hot
bubble, as well as from additional absorption of the distant component.

\section{Conclusion}

We have used a set of deep observations with the \sax\ LECS
to study the high galactic latitude CXB spectrum.
We show that the spectrum is consistent with both a
power-law plus two thermal components model (as currently accepted for
the ROSAT PSPC plus ASCA SIS background determinations), and with a
broken power-law plus single thermal component model. 

While the power-law parameters are in good agreement with the
ASCA SIS measurements, the temperatures of the thermal components are
significantly different from those obtained with the PSPC.
In particular, the ``super-soft'' thermal component seen
by the PSPC is not evident in the LECS spectrum.

\begin{acknowledgements}
  We thank G. Cusumano for providing the
  results of the in-flight vignetting calibration, F.~Fiore and P.~Giommi
  for discussions, and S.~Molendi for the comparison with 
  MECS results. The \sax\ 
  satellite is a joint Italian-Dutch programme. M.~Guainazzi,
  T.~Oosterbroek, and A.~Orr acknowledge ESA Fellowships. 
\end{acknowledgements}

\end{document}